\documentclass[twocolumn,showpacs,preprintnumbers,amsmath,amssymb,floatfix]{revtex4}

\usepackage{graphicx}
\usepackage{dcolumn}
\usepackage{bm}

\begin{document}

\preprint{APS/123-QED}

\title{Multiple Timescale Dynamics in Economic Production Networks}
\author{Adam Ponzi}
\email{adam@complex.c.u-tokyo.ac.jp}
\author{Ayumu Yasutomi}
\author{Kunihiko Kaneko}%
\affiliation{College of Arts and Sciences, Tokyo University, Meguro-ku, Tokyo, Japan}
\date{\today}
             
\begin{abstract}
We describe a new complex system model of an evolving production
economy. This model is the simplest we can envisage which incorporates
the new observation that the rate of an economic production process
depends only on the {\it minimum} of its supplies of inputs. We
describe how this condition gives rise to a new type of complex
multiple timescale dynamical evolution through a novel type of
bifurcation we call a {\it trapping bifurcation}, which is also shown
to be one cause of non-equilibrium economic behaviour. Such dynamics
is an example of {\it meta-level} coupling which may also arise in
other fields such as cellular organization as a network
of molecular machines.
\end{abstract}

\maketitle
\underline{Introduction.} 
Recently physicists have been devoting a great deal of attention to
economic and financial phenomena\cite{mant}. Economic dynamics is
easily observed to be far from equilibrium where periodic recessions,
unemployment and unstable prices occur persistently. An understanding
of the origins of this behaviour from the viewpoint of complex
dynamical systems theory would be very valuable. Our
model\cite{physA}, which we hope takes a step in this direction, is
based on von-Neumann's\cite{vmpap-vmbook} neoclassical model of
economic production and catalytic chemical reaction network
dynamics\cite{kan}.

\begin{figure}
\centerline{
{\resizebox{0.48\textwidth}{30mm}
{\includegraphics{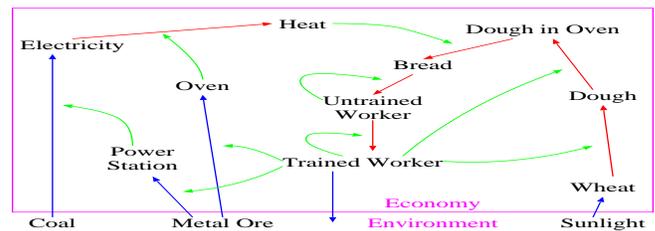}}}}
\caption{An Economy. Red lines are production flows, blue lines flows to and from the environment (including other countries) and green lines are catalytic effects. Some processes are autocatalytic.}
\label{econ}
\end{figure}

The original von Neumann model (VNM) of economic production assumes
that each good is produced jointly with certain others, in an
analogous way to a chemical reaction. A production process is the
operation which converts one bundle of goods, including capital
equipment, into another bundle of goods, including the capital
equipment. Capital goods therefore function approximately like {\it
catalysts} in chemical reactions, reformed at the end of the reaction
in amounts conserved in the reaction. Consumption of goods is assumed
to take place only through production processes, including life
necessities consumed by workers, and all income is reinvested in
production. Therefore the VNM is defined by a fixed input matrix and a
fixed output matrix, representing, the fixed stochiometric ratios of
input products and output products for each process. An example of
such an economic network is shown in Fig.\ref{econ}. Each process is
assumed to have unit time duration, longer processes being broken down
into several processes with intermediate products. The VNM is defined
as a static equilibrium model describing relationships between the
variables which must hold at equilibrium. Equilibrium is a state of
`balanced growth' where prices are constant. There are no dynamics
defined by the model which might describe out of equilibrium or
approach to equilibrium behaviour.

However it is rarely the case that economic processes are in
equilibrium. 
For example consider a simple bakery with input: (dough, baker, oven)
and output: (bread, baker, oven), so that the baker and the oven are
catalysts. We observe that the {\it rate} of bread production, and
economic processes in general, depends on the quantity of the {\it
minimum} of its input supplies and not on the quantities of its other
supplies. I.e. the baker working at full pace can only fill ovens at a
certain rate. Similarly employing more bakers will not increase bread
production if the ovens are already full. This is in direct contrast
with a chemical reaction obeying the law of mass action, where an
increase in concentration of any of its input species will increase
the rate of reaction.

This minimum condition leads directly to {\it multiple
timescales}\cite{fuj}. Consider a factory assembling bicycles from
wheels and frames at a given rate. Suppose for some reason the rate of
supply of wheels increases so that their price drops. To maximise
efficiency the factory will try to use all its funds to expand
production, buying the extra wheels and demanding more frames. A
factory with large funds may demand frames so strongly that the the
frame-makers respond to it and increase their supply rate. In this way
the wheel-makers and frame-makers production rates may become
synchronised. A factory with small funds on the other hand may only
weakly influence the frame supply rate. Its bicycle production rate
will be determined by the supply rate of whichever is the most
expensive, and may {\it switch} between the rates as price levels
change. The factory will have inefficient periods with surpluses of
frames or wheels, but to avoid this by not using all its funds would still
not be maximally efficient.

Here we show how such multiple timescales arise `endogenously' in a
new model of an economy of coupled processes which is different from
the VNM in that we explicitly take into account the way a process'
production rate depends on the minimum of its input supplies.

\underline{Model.} 
Although our model ignores many details, we believe it captures the
essential characteristics of a evolving production-marketing network,
in the simplest way possible. The system is defined by a fixed number
of processes $K_I$ and products $K_J$ with input stochiometric ratios
$a_{ij}$ and ouput stochiometric ratios $b_{ij}$, where $i$ labels the
processes and $j$ the products. Each process has {\it supplies} of its
products $S_{ij}(t)$, it also has a dynamical fitness called {\it
funds} $F_i(t)$, measured in dollars, which represents the `size' or
`intensity' of the process. The model can be conveniently broken down
into four simple parts. 

(i) {\it Processing.} Here all processes
simultaneously manufacture goods so that their supplies after
processing, denoted with `$^*$', are given by,
\begin{equation}
S^*_{ij}(t)=S_{ij}(t)+(b_{ij}-a_{ij}) Min_k(S_{ik}(t)/a_{ik})\label{s*}
\end{equation}
where the processing rate is given by
$Min_k(S_{ik}(t)/a_{ik})$ where $Min_k(x_k)$ denotes the minimum over
$x_k$. 

(ii) {\it Product revaluation}. The {\it value}, or price
$p_j(t)$, of a product is recalculated after processing and depends on
the overall supply available in the whole economy and on how much all
the processes want it. To measure how much the processes want the
product we consider how much they will pay for it and each process
therefore divides its funds $F_i(t)$ into {\it demands} $D_{ij}(t)$
for its input products. The new product value $p_j(t)$ is then given
by
\begin{eqnarray}
p_j(t)=\frac{Total\hspace{0.1cm}Demand}{Total\hspace{0.1cm}Supply}=\frac{\sum_i D_{ij}(t)+D^{e}_j}{\sum_i S^*_{ij}(t)+S^{e}_j},\label{pri}
\end{eqnarray} 
where $S^{e}_j$ and $D^{e}_j$ are possible external supplies and
demands coming from, say, another country. 

There are many ways a process may allocate its funds into demands. If
we consider a feedback from the price the funds should be divided
according to $a_{ij}p_j(t-1)$ since this is process $i$'s best
estimate of the per unit cost of input $j$ depending on the most
recent known price $p_j(t-1)$. Alternatively the process may simply
divide its funds according to the ratios $a_{ij}$. Here we will
consider,
\begin{equation}
D_{ij}(t)=F_{i}(t)(\gamma \frac{a_{ij}p_j(t-1)}{\sum_ja_{ij}p_j(t-1)}+(1-\gamma)a_{ij})\label{dem}
\end{equation}
where the parameter $0<\gamma<1$ quantifies the feedback from the
price. In fact we will show that the interesting behaviour is found
independently of $\gamma$, the important points being
$D_{ij}(t)\propto F_i(t)$ $\forall j$ and $\sum_j D_{ij}(t)=F_i(t)$.

(iii) {\it Product resharing}. Processes obtain new
supplies for processing simply according to
\begin{equation}
S_{ij}(t+1)=D_{ij}(t)/p_j(t)\label{sup}
\end{equation} 
so that processes with relatively large funds obtain larger
proportions of the available supply. 

(iv) {\it Process
revaluation}. The new process funds are,
\begin{equation}
F_i(t+1)=\sum_j S^*_{ij}(t)p_j(t)\label{fit}
\end{equation} 
so that processes with larger manufactured supplies obtain greater
proportions of the available funds, and in this way the system becomes
an evolving network. 

\begin{figure}
\centerline{
{\resizebox{0.48\textwidth}{70mm}
{\includegraphics{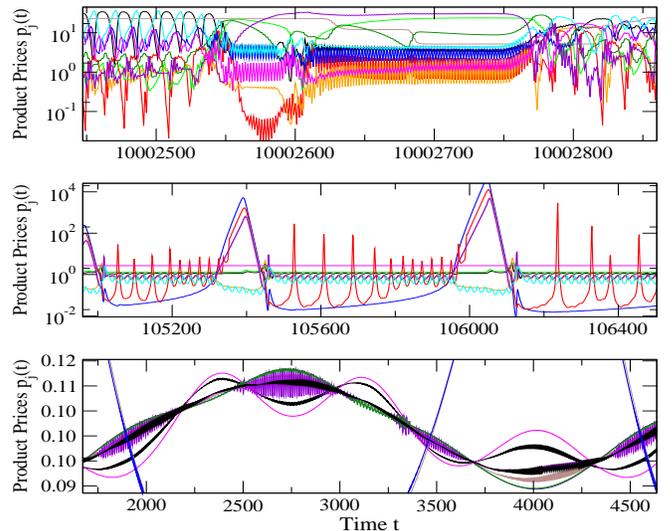}}}}
\caption{
Three examples of multi-timescale time series produced from 15
process, 10 product networks, ($\gamma=0.9$).  }
\label{multi}
\end{figure}

\underline{Results.} These equations have the novel property that they {\it
switch} their form depending on which product is the minimum at any
time. Synergetic effects then appear because the dynamics itself
selects the equations which determine its own future evolution. We
will show that this strong non-linearity leads to the new type of very
complex multi-timescale dynamics illustrated in Fig.\ref{multi},
focusing on $\gamma=0$ since the analysis is much simpler.

We first show that even the simplest single process, 
in a fixed environment, will, under general conditions,
oscillate periodically due to a novel switching mechanism and
furthermore this has interesting economic relevance. 

This $K_I=1$, $K_J=3$ system has one input product I, one catalyst C,
and one output product O, reacting with non-zero stochiometric ratios
$a_I=a_C=b_C=b_O=1/2$ with $\gamma=0$. It has 3 variables, its input
supplies $S_I(t)$, catalyst supplies $S_C(t)$ and funds $F(t)$, where
we drop the process label $i$. There are two {\it phases}: (X) excess
input, $S_I(t)>S_C(t)$ and (Y) excess catalyst, $S_I(t)<S_C(t)$, such
that Eq.\ref{sup}, $S_I(t+1)=1/2F(t)/p_I(t)$ and
$S_C(t+1)=1/2F(t)/p_C(t)$ becomes,
\begin{eqnarray}
(X)\hspace{0.5cm}S_I(t+1)=\frac{1/2F(t)(S_I(t)-S_C(t)+S^{e}_I)}{1/2F(t)+D^{e}_I}\label{dyn1}\\
(Y)\hspace{0.5cm}S_I(t+1)=\frac{1/2F(t)S^{e}_I}{1/2F(t)+D^{e}_I}\label{dyn2}\\
(X,Y)\hspace{0.5cm}S_C(t+1)=\frac{1/2F(t)(S_C(t)+S^{e}_C)}{1/2F(t)+D^{e}_C}\label{dyn3}
\end{eqnarray}
where we have omitted the (X) and (Y) $F(t)$ equations for simplicity
and where $S^{e}_I,S^{e}_C,S^{e}_O$ and $D^{e}_I,D^{e}_C,D^{e}_O$ are
fixed external supplies and demands of the three products
respectively. Here, since $\gamma=0$ and $a_I=a_C$, (X) and (Y) {\it
switch} whenever the prices $p_I(t)$ and $p_C(t)$ cross.

To understand the novel dynamical features of this system it is
sufficient to consider the interaction of the phase (X) and phase (Y)
fixed points. Denoting these two fixed points by superscript
`$^{x,y}$' we find that, although the fixed point positions are
different, the fixed point prices in both phases are the same and
given by,
\begin{equation}
p_I^{x,y}=p_O^{x,y}=\frac{D^e_I+D^e_O}{S^e_I+S^e_O},\hspace{1.0cm}p_C^{x,y}=\frac{D^e_C}{S^e_C}.
\label{extpri}
\end{equation}
Therefore a process in equilibrium acts to make its input and output
product prices equal\cite{effic}. This interesting non-trivial
relationship has the observable consequence that for example scaling
both $D^e_O$ and $S^e_O$ by the same constant factor will change the
price of both $O$ and $I$, although one would naively expect that such
a scaling should leave the prices unchanged.\cite{zerofp}
\begin{figure}
\centerline{
{\resizebox{0.48\textwidth}{60mm}
{\includegraphics{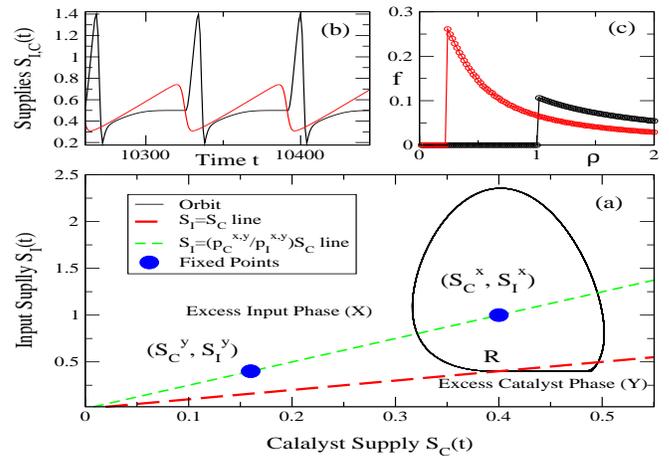}}}}
\caption{\label{orbit}
(a) The {\it switching state}; creation of periodic orbit from node
and focus. The $S_I=S_C$ line (see key) divides the (X) and (Y)
phases, whose fixed points are shown labelled. The fixed points are
such that they lie on the $S_I=S_C(p_C^{x,y}/p_I^{x,y})$ line, (see
key). Therefore only one can be {\it compatible} with its phase. Here,
in the switching state shown, where $p_C^{x,y}/p_I^{x,y})>1$ the (X)
fixed point is compatible since $S_I^x>S_C^x$, but the (Y) fixed point
is incompatible since $S_I^y>S_C^y$. When the trajectory is in the (Y)
phase it is attracted to the stable node $(S_C^y,S_I^y)$, which is
incompatible with phase (Y). Therefore the phase boundary must be
crossed and the system switches to an oscillation around the
$(S_C^x,S_I^x)$ focus. This then causes the system to be reinjected
into the (Y) phase again, producing the clockwise orbit. Furthermore,
regardless of initial conditions, the system is always reinjected at
the point $R$ thereby producing a limit cycle. Since $S_I^y=S_C^x$,
the fixed points approach each other as $p_C^{x,y}/p_I^{x,y}$
approaches 1 from above and therefore the {\it switching frequency}
increases while the amplitude decreases as confirmed in (c). (b) Time
series of input supply $S_I(t)$ (black) and catalyst $S_C(t)$ (red)
for a single process with $\gamma=1$. (c) Variation of switching
frequency $f$ against $\rho=(p_C^{x,y}/p_I^{x,y})$ from
Eq.\ref{extpri}. $\gamma=0$ (black), $\gamma=1$ (red). As $\gamma$
increases the bifurcation point decreases and the frequency $f$
increases.}
\end{figure}

However as shown in Fig.\ref{orbit}(a), which shows a typical time
series variation in the $S_I(t)$, $S_C(t)$ plane, the process will not
in general be at a fixed point, but oscillates periodically. Although
neither phase (X) nor (Y) has a periodic attractor on its own, a limit
cycle is created by a novel trapped switching mechanism which arises
due to the minimum dynamics as explained in Fig.\ref{orbit}(a). As
shown the {\it ratio of the fixed point values}
$\rho=(p_C^{x,y}/p_I^{x,y})$ acts as a {\it trapping bifurcation}
parameter since when $\rho>1$ the switching state is obtained but when
$\rho<1$ the system remains at the node without oscillating.

The switching state is plausible economically. In the switching state
we find the fund fixed points are such that $F^x=\rho
F^y>F^y$. Therefore when the system is in the excess catalyst phase
(Y) its funds $F(t)$ decrease and it acts to reduce $S_C(t)$
(Eq.\ref{dyn3}). However the decrease in $S_C(t)$, since it is not the
minimum supply, has no effect on the processing rate, and $S_I(t)$
(Eq.\ref{dyn2}) is therefore not affected by the change in
$S_C(t)$. In the excess supply phase (X) on the other hand, the funds
$F(t)$ increase and the process acts to increase $S_C(t)$
(Eq.\ref{dyn3}) which increases the processing rate and decreases
$S_I(t)$ (Eq.\ref{dyn1}) producing the oscillation. Furthermore as
explained in Fig.\ref{orbit}(a)(c) the switching frequency decreases
as $\rho$ increases. This is to be expected economically. The price
fixed points are determined by the external environment. When
$p_C^{x,y}>>p_I^{x,y}$ the process demands will have much less affect
on $p_C(t)$ than $p_I(t)$. Therefore in the excess supply phase (X)
$S_C(t)$, and therefore the processing rate, will only increase slowly
and $S_I(t)$ will be reduced only slowly, producing large amplitude
slow oscillations. 

We therefore expect that even a single process in a fixed environment
will have periodic dynamics, with periods of (X) full employment
($S_I(t)>S_C(t)$) interrupted by periods of (Y) unemployment,
($S_I(t)<S_C(t)$).

Through simulations we find Eqs.\ref{extpri} also seem to hold in the
$\gamma>0$ case. The same switching state appears, see
Fig.\ref{orbit}(b)(c), but the exact bifurcation point
varies. Furthermore the equivalence between switching and price
crossing, seems to be still approximately true. However the switching
dynamics is more complex as shown in Fig.\ref{orbit}(b). This is
interesting since one would naively expect that using price
information should allow the process to obtain more balanced input
supplies and processing to occur more smoothly. However doing so moves
the process nearer the trapping birfurcation enhancing the switching
feedback effects. This can produce quasiperiodic or weakly chaotic
dynamics.
\begin{figure}
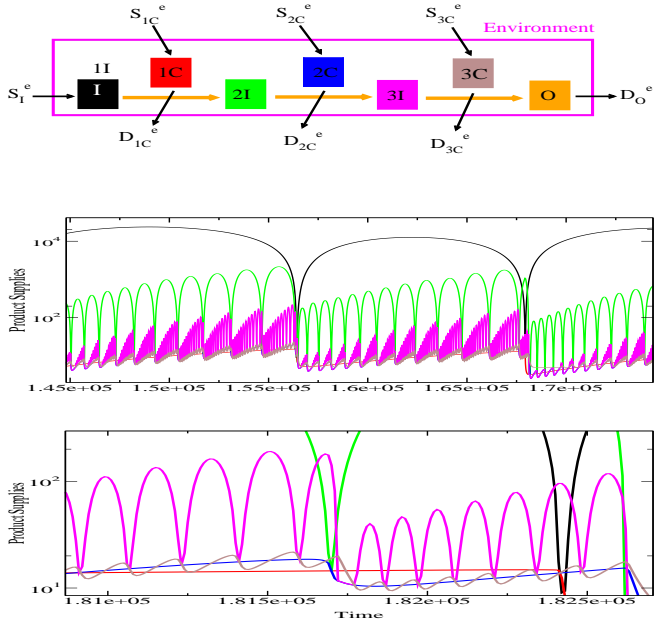

\centerline{
{\resizebox{0.48\textwidth}{20mm}
{\includegraphics{fig4.eps}}}}\vspace{0.8cm}
\centerline{
{\resizebox{0.48\textwidth}{25mm}
{\includegraphics{fig5.eps}}}}\vspace{0.3cm}
\centerline{
{\resizebox{0.48\textwidth}{25mm}
{\includegraphics{fig6.eps}}}}
\caption{
{\it Hierarchical multiple timescale limit cycle attractor} in process
chain in fixed environment. (a) Diagram of a 3 process 7 product
chain. 3 products are catalysts, one in each process each with a fixed
external supply $S_{iC}^{e}$ and demand $D_{iC}^{e}$, defining an
external price $p_{iC}^{x,y}=D_{1C}^{e}/S_{1C}^{e}$ for each
catalyst. There is one input product $I$ with a fixed external supply
$S_{I}^{e}$ and one output product $O$ with fixed external demand
$D_{O}^{e}$. There are two intermediate products with no external
supplies and demands. (b) Time series of supplies and detail (c),
colours are as depicted in (a). Due to price equalization
Eq.\ref{extpri} in chains of processes we get
$p_{I}^{x,y}=p_{O}^{x,y}=D_{O}^{e}/S_{I}^{e}$ for the input and output
prices. Here we set external environment such that
$p_{1C}^{x,y}>>p_{2C}^{x,y}>>p_{3C}^{x,y}>p_{I}^{x,y}$ so that the 3
processes are all in the trapped state with very different switching
frequencies, and we obtain a multiple timescale limit cycle. Since the
3 processes are in the same chain and they are all in catalyst minimum
state the process 3 brown catalyst oscillates around the output level
of process 2 which is the blue catalyst which oscillates around the
the output level of process 1 which is the red catalyst, to give the
pleasing (but economically undesirable) effect of sewing. Here price
feedback $\gamma=0.9$.}
\label{gam}
\end{figure}

To understand the behaviour of a multi-process economy and the
appearance of complex multi-timescale dynamics, shown in
Fig.\ref{multi}, we need to see how individual processes affect each
other and in this respect we need two points. (i) {\it Trapping
bifurcation produces extremely different frequencies.}  When there are
several processes each process has its own, now time dependent,
trapping bifurcation parameter $\rho^i(t)=p_C^i(t)/p_I^{i}(t)$ and own
switching frequency, determined by its external environment, where the
superscript $^i$ now refers to each process' `fixed point' prices
given by Eq.\ref{extpri}. The switching state can be turned on and off
when the fixed point prices $p_C^i(t)$ and $p_I^{i}(t)$ cross and as
explained above, since the switching frequency depends strongly on
$\rho$, not on the complex eigenvalues of focus in the conventional
way, switching frequencies can be extremely different.

(ii) {\it Unequal coupling and transfer of oscillations.} We note that
this system is an evolving system of competing coupled oscillators
that can have very different funds $F_i(t)$. They are coupled only
through a {\it series} of prices (Eqs.\ref{pri}-\ref{fit}), so that not
all processes are directly coupled. Furthermore the price couplings
(Eq.\ref{pri}) are of mean-field {\it threshold} type, so that each
process transfers it oscillations with different `weights' determined
by their different funds $F_{i}(t)$ which control the size of their
demands $D_{ij}(t)$, (Eq.\ref{dem}). Due to this the oscillations of a
given process may dominate its input product price for example but
hardly appear in its catalyst price, allowing multiple timescales and
indeed trapping itself to appear.

An example of how these (i) and (ii) combine in a multi-process system
is the simple chain shown in Fig.\ref{gam}.

\underline{Summary.}
It is clear that when economic processes with minimum condition
dynamics are coupled together, price level crossings will induce
hierarchical trapping bifurcations with different switching
frequencies in individual processes. These oscillations will be
transmitted to some prices but not others and the dynamics can easily
become very complex as illustrated in Fig.\ref{multi} which was
produced from random networks similar to that shown in
Fig.\ref{econ}. This behaviour may be the fundamental origin of
persistent cyclic non-equilibrium economic dynamics.

The novel trapping bifurcation and switching state dynamics introduced
here may be relevant in other fields with meta-level coupling where
real variable dynamics is synergetically mixed with boolean type
logic. Economic agents {\it choosing} a {\it maximum} utility action,
cellular reaction path expression interacting with genetic {\it
sorting}, networks of {\it molecular machines} in a cell and neural {\it
decision} making systems are possible examples.

We thank K.Fujimoto for useful discussions. A.Ponzi wishes to
thank the Japan Society for Promotion of Science for support of this
work.

\begin{thebibliography}{999}
\bibitem{effic}
This relationship is only approximate when processes are less than
100\% efficient or profit taking occurs. In this case input
prices must be less than output prices.
\bibitem{zerofp}
In fact each phase (x) and (y) also has another fixed point at the
origin, but in each phase only one of the two fixed points is
stable. In fact in each phase the zero fixed point is stable and the
interior fixed point unstable when $(D^e_I/S^e_I)>(D^e_O/S^e_O)$ and,
as is plausible economically, the process goes bankrupt.
\end{thebibliography}

 \end{document}